\documentclass[journal]{IEEEtran}
\usepackage{fontenc}
\usepackage{multirow}
\usepackage{cellspace}
\usepackage{caption}
\usepackage{stfloats}
\usepackage{makecell}
\usepackage[colorlinks,
linkcolor=black,
anchorcolor=black,
citecolor=black]{hyperref}
\usepackage{amsfonts}
\usepackage{mathrsfs}
\usepackage{amsfonts}
\usepackage{ifpdf}
\usepackage[justification=centering]{caption}
\usepackage{cite}
\ifCLASSINFOpdf
\usepackage[pdftex]{graphicx}
\else \fi
\usepackage[cmex10]{amsmath}
\usepackage{amsmath,amsfonts}
\usepackage{array}
\usepackage{mdwmath}
\usepackage{mdwtab}
\usepackage{eqparbox}
\usepackage[tight,footnotesize]{subfigure}
\usepackage{algorithm}
\usepackage{algorithmicx}
\usepackage{algpseudocode}
\usepackage{amsmath}
\usepackage{epsfig}
\usepackage{ae}
\usepackage{amssymb}
\usepackage{times}  
\usepackage{booktabs}
\usepackage{color}
\usepackage{bm}

\usepackage{epstopdf}
\usepackage{stfloats}
\usepackage{subfigure}
\usepackage{float}
\usepackage{url}
\usepackage{threeparttable}

\floatname{algorithm}{Algorithm}

\begin{document}
\title{Network Intent Decomposition and Optimization for Energy-Aware Radio Access Network}

\author{Yao~Wang,~\IEEEmembership{Graduate Student Member,~IEEE,}
	Yijun~Yu,
	Yexing~Li,
	Dong~Li,
	Xiaoxue~Zhao,
	and Chungang~Yang,~\IEEEmembership{Senior Member,~IEEE}
	\vspace{-1.5em}
	\thanks{This work was supported by a cooperative project from Huawei Technologies Company Ltd. (TC20221114023).}
	\thanks{Yao Wang, Xiaoxue Zhao, and Chungang Yang are with the State Key Laboratory of Integrated Services Networks, Xidian University, Xi'an, 710071, China (email: yaow0518@gmail.com; chgyang2010@163.com; zhaoxiaoxue1123@163.com).}
	\thanks{Yijun Yu is with the Research Department of Autonomous Driving Network, Huawei Technologies Company Ltd., Dongguan, 523808, China (e-mail: yuyijun@huawei.com).}
	\thanks{Yexing Li, and Dong Li are with the Research Department of Autonomous Driving Network, Huawei Technologies Company Ltd., Shanghai, 201206, China (e-mail: liyexing@huawei.com; layton.lidong@huawei.com).}
}\maketitle

\begin{abstract}
With recent advancements in the sixth generation (6G) communication technologies, more vertical industries have encountered diverse network services. How to reduce energy consumption is critical to meet the expectation of the quality of diverse network services. In particular, the number of base stations in 6G is huge with coupled adjustable network parameters. However, the problem is complex with multiple network objectives and parameters. Network intents are difficult to map to individual network elements and require enhanced automation capabilities. In this paper, we present a network intent decomposition and optimization mechanism in an energy-aware radio access network scenario. By characterizing the intent ontology with a standard template, we present a generic network intent representation framework. Then we propose a novel intent modeling method using Knowledge Acquisition in automated Specification language, which can model the network ontology. To clarify the number and types of network objectives and energy-saving operations, we develop a Softgoal Interdependency Graph-based network intent decomposition model, and thus, a network intent decomposition algorithm is presented. Simulation results demonstrate that the proposed algorithm outperforms without conflict analysis in intent decomposition time. Moreover, we design a deep Q-network-assisted intent optimization scheme to validate the performance gain.
		
\end{abstract}
	

\section{Introduction}
\IEEEPARstart{W}{ith} recent advancements in the sixth generation (6G) communication technologies, more vertical industries have encountered diverse network services. Radio access network (RAN) empowered by 6G provides user equipment (UE) with higher throughput, lower latency, and quality-assured wireless communication services \cite{LinKH2023,PoleseM2024}. However, the huge number of base station (BS) accesses requires huge bandwidth, and generates very high energy consumption \cite{LeeYL2021}. How to reduce energy consumption is critical to meet the expectation of the quality of diverse network services. There have been many efforts from academic and standardization fields on the RAN, where the design of network architecture and functionality \cite{CoronadoE2022}, and the definition of network properties \cite{Standard2,Standard3}, while improving network performance and evaluating energy consumption.

Due to the emergence of diverse network services in 6G vertical industries, the RAN is facing exponential growth in service demands. As the number of user accesses grows, there are problems such as more complex network architecture, higher dynamics, and highly constrained resources also become more apparent. Moreover, the size of the network and the types and number of BSs and adjustable network parameters are becoming increasingly large and coupled to each other. The above problems create new demands and challenges for the continued development of the RAN. Meanwhile, the network management, configuration, and execution of the RAN are operated by domain experts \cite{BenzaidC2020}, which has resulted in the coupling of complex manual operations, high operation and maintenance costs, and limited network services. Therefore, existing network architectures can no longer meet the precise, real-time analysis for service requirements, and the automation capability of the network needs to be improved.

To achieve a higher level of network automation capability, network intent representing service requirements needs to be modeled, decomposed, and optimized. From the perspective of network intent decomposition and optimization reasonability, when the BSs are required to fulfill network intent related to energy-saving, the operator as a manager that needs to adjust energy-saving operations. The objective of the operator is to analyze the status of network objectives and determine what types of energy-saving operations will meet network intent. Such analysis motivates us to  better understand the interaction between the BSs and the UEs. Aiming to address network intent decomposition and optimization concerns, there are three key bottlenecks that must be overcome: $i)$ how to accurately represent the intent and network ontologies of network intent, $ii)$ how to systematically map energy-saving operations according to the status of network objectives, and $iii)$ how to select the optimal energy-saving operation for the BSs through intelligent decision-making methods. These bottlenecks and challenges motivate the need for a better network intent decomposition and optimization scheme design. 

To our best knowledge, existing works do not involve network intent modeling, decomposition, and optimization for the RAN, and this is the first work. Compared to conventional optimization methods, we can find that the exploration of network intent decomposition and optimization has become highly valuable. For bridging the research gap, in this paper, we achieve accurate network intent decomposition and effective network objective optimization by identifying such an interaction between the BSs and the UEs. The main contributions of this paper can be summarized as follows.

\begin{itemize}
	\item 
	We present a 3GPP template-based network intent representation method to characterize the intent ontologies.
	
	\item 
    We identify the elements of the network ontology in network intent modeling process, and utilize a \textit{Knowledge} Acquisition in automated Specification language to describe these elements.
	
	\item 
	We introduce a \textit{Softgoal} Interdependency Graph decomposition mechanism to decompose the network intent for achieving the logic of ``network intent''--``network objectives''--``energy-saving operations''.
	
	\item 
	We design a network intent decomposition algorithm, and validate the initial decomposition performance. Moreover, a deep Q-network-assisted intent optimization scheme is presented and compared with the baselines.
\end{itemize}

The subsequent sections of this paper are structured in the following manner. In Section \ref{sectionII}, we provide an overview of the related works. Section \ref{sectionIII} provides an explanation of the system model. In Section \ref{sectionIV}, the network intent representation, transformation and modeling method is proposed. Section \ref{Section-V} presents the network intent decomposition scheme. The presentation of the simulation results is found in Section \ref{sectionVI}. Ultimately, this work is concluded in Section \ref{sectionVII}.

\section{Related Works}
\label{sectionII}

\subsection{Representation Models for Network Intent}
Recently, the integration of intent representation into mobile networks has received considerable attention due to the accuracy of intent representation can be enhanced by jointly modeling $expectationObjects$ and $expectationTargets$ of the intent \cite{StandardIntent}. In \cite{CohenR2013}, the authors explored a novel intent modeling approach for abstracting network services focusing on identifying network requirements. The authors in \cite{HeavenW2004} developed a construction framework for the goal model to learn and refine via jointly considering temporal linear logic and the connections among goals. Similar to \cite{HeavenW2004}, the authors of \cite{MateA2014} proposed an $i^*$ framework to extend the $i^*$ profile to model user requirements by adding the capability of error correction. Compared with the studies applying \textit{Knowledge} Acquisition in automated Specification (KAOS) and $i^*$ approach for goal modeling, the authors in \cite{MylopoulosJ1992} proposed a softgoal-oriented NFR framework to demonstrate the function of goal reasoning. The authors in \cite{NguyenCM2016} presented an extended modeling language called constrained goal models to express the preferences among the goals and define the constraints related to the current environment. Ultimately, these works mainly focused on software requirements analysis \cite{HeavenW2004,MateA2014,MylopoulosJ1992} to model the requirements of users to conform to a generic system description.

\subsection{Decomposition Techniques for Network Intent}
Several recent works have been devoted to dividing the technique of intent decomposition in communication systems into two types, e.g., architecture-based \cite{NguyenCD2009,LesireC2022} and methodology-based \cite{ScheidEJ2017,LeivadeasA2021,BonfimM2021}. For multi-agent systems, the object-oriented structures were presented in \cite{NguyenCD2009} by mapping the agent-oriented abstractions to test whether the goals are met. In \cite{LesireC2022}, the authors proposed a hierarchical deliberative framework and realized the management of task specifications with multiple goals and tasks. For methodology-based intent decomposition, the authors in \cite{ScheidEJ2017} proposed an integrated network function virtualization (NFV)-based intent refinement solution for intent-driven networks to refine the intents into a set of configurations. Similar to \cite{ScheidEJ2017}, the authors in \cite{LeivadeasA2021} proposed an automatic intent-based network-based virtual network function (VNF) deployment solution to refine abstract requests into VNF deployment policies. In the same scenario, the authors of \cite{BonfimM2021} developed an Onto-Planner semantic model to refine the management and orchestration tasks. In summary, the above studies introduced the intent decomposition process, and captured the important effect of formed sub-goals and operations on the system.

\subsection{Optimization Methods for Network Intent}
Recently, several researches have been attempted to deal with objective optimization problems in wireless networks using different methodologies and algorithms. To predict environmental conditions in advance, the authors in \cite{GrassiV2009} introduced a self-adaptive planning mechanism to verify the quality of service. In order to accomplish resource management in heterogeneous cellular networks, article \cite{LiuL2021} developed a gravitational search algorithm-based multi-objective optimization approach to reduce grid energy consumption. The establishment time of collaborative beamforming and the energy simultaneous minimization problem was presented in \cite {SunG2021} via optimizing excitation current weights, locations, and flight speeds for unmanned aerial vehicle networking. In the same network architecture for \cite{SunG2021}, the authors in \cite{YuY2021} built a 3-tuple objective optimization problem via jointly optimizing energy consumption, data rate, and harvested energy. The authors of \cite{GaoX2023} formulated an ultra-reliable low latency communication-oriented multi-objective optimization problem. Finally, network intent can be accurately converted into a multi-objective optimization problem. The research efforts devoted to objective optimization methods and energy consumption have also become important optimization objectives \cite{SunG2021,YuY2021}.

\section{System Model \& Problem Formulation}
\label{sectionIII}
\subsection{System Overview}
\begin{figure}[t]
	\centering
	\includegraphics[width=2.8in]{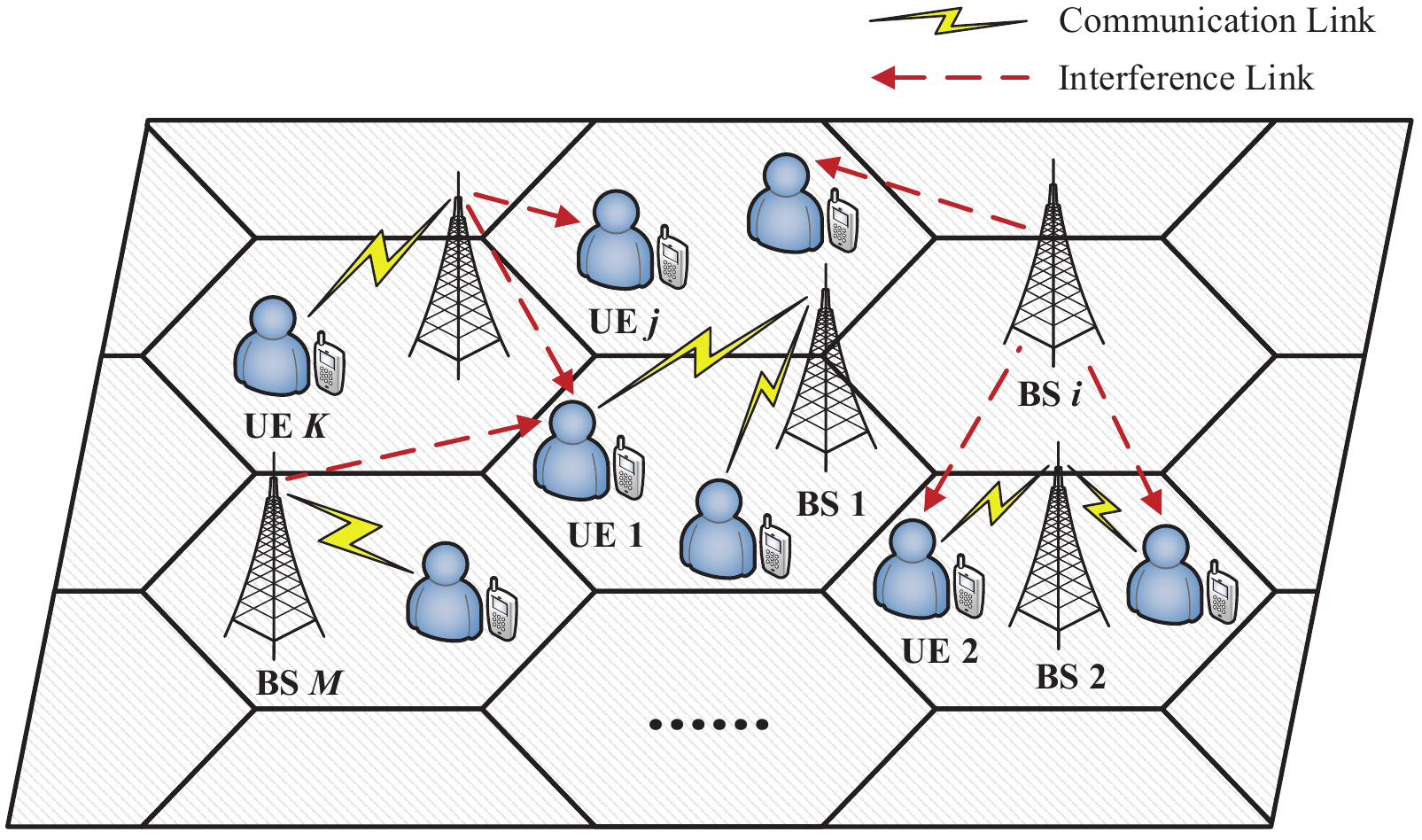}
	\caption{The scenario of energy-aware radio access network.}
	\label{Fig_1_Scenario}
	\vspace{-0.5cm}
\end{figure}

 We consider an energy-aware RAN scenario, as illustrated in Fig. \ref{Fig_1_Scenario}, consisting of $M$ BSs and $K$ UEs at time $t$. The set of BSs and UEs are denoted as ${\cal M} = \left\{ {1,2, \cdots, M} \right\}$, $\forall i \in {\cal M}$ and ${\cal K} = \left\{ {1,2, \cdots, K} \right\}$, $\forall j \in {\cal K}$, respectively. Each BS is equipped with $M_{\rm T}$ transmission antennas to coverage the UEs equipped with $K_{\rm R}$ receiving antennas. Each BS is operating in a fixed altitude of $L_{\rm B}$. Here, we assume that the UEs move continuously, and that the speed at which the UEs move follows a Gaussian distribution $|{\cal N}(\mu, \sigma^2)|{\rm m{s^{ -1}}}$ of mean $\mu$ and variance $\sigma^2$. The UEs can randomly choose the direction to move. Without loss of generality, the location of each communication entity is depicted using a 3D Cartesian coordinate system. The location of BS $i$ at time $t$ is specified by $\left( {x_i^{},y_i^{}} \right) \in {\mathbb{R}}^{2 \times 1}$. Similarly, the location of UE $j$ at time $t$ can be represented as $\left( {x_j^{},y_j^{}} \right) \in {\mathbb{R}}^{2 \times 1}$. 

To establish the downlink between the BSs and the UEs, it is assumed that the BSs through the communication mode of orthogonal frequency division multiplexing (OFDM) to schedule the access of multiple UEs. The OFDM mode is to partition overall system radio resources into physical resource blocks (RBs) from time and frequency domains. When the system bandwidth of BS $i$ is $B_i$, the number of physical RBs allocated to BS $i$ is $r_i=B_i/B^{\rm RB}$, where $B^{\rm RB}$ is a constant, and denoted as the bandwidth of each RB. The UEs select the BSs, and then add downlink communication services to the buffer of the BSs. The service flow of UE $j$ is abided by a Poisson distribution $X \sim P\left(\chi_j \right)$ of service arrival rate $\chi_j$. The total packet size of UE $j$ at time $t$ is defined by $W_j$ (in bits). We assume that the time interval among the arrival of neighboring packets $T_j$ obeys an Exponential distribution $f(T_j^{})$ with an expectation value $1/\chi_j$. In each transmission time interval (TTI), the BSs assign the RBs to perform resource scheduling for meeting the service requirements of the UEs. The UEs map the downlink channel quality indicator (CQI), and report it to the BSs. The set of CQI numbers is represented as ${\cal N} = \left\{ {1,2, \cdots, N} \right\}$. After receiving CQI $n \in {\cal N}$ reported by UE $j$, BS $i$ determines the modulation and coding scheme.

\subsection{Energy Consumption Model}
The success of energy-saving operations is mostly reflected in the energy consumption profile. We consider that there are two components to the total energy consumption: the energy consumption proportional to the utilization of the RBs and fixed energy consumption \cite{SonK2011}. As such, to illustrate the connection relationship between BS $i$ and UE $j$ at time $t$, a binary variable is defined

\vspace{-0.25cm}
\begin{equation}
	{\xi _{i,j}} = \left\{ {\begin{array}{*{20}{l}}
			{\!\!\!1,}&{{\text{if UE }}j{\text{ is connected to BS }}i,}\\
			{\!\!\!0,}&{{\text{otherwise}}{\rm{.}}}
	\end{array}} \right.
\end{equation}
\vspace{-0.25cm}

We define the load $\theta_i$ as the ratio of the number of RBs that have been utilized $\sum\nolimits_{j \in {{\cal K}}} {{\xi _{i,j}}} {r_{i,j}}$ to the total number of RBs $r_i$, the load of BS $i$ at time $t$ can be given by

\vspace{-0.2cm}
\begin{equation}
{\theta _i} = \frac{{\sum\nolimits_{j \in {{\cal K}}} {{\xi _{i,j}}} {r_{i,j}}}}{{{r_i}}}, \:\: \forall i \in {{\cal M}},\forall j \in {{\cal K}},
\end{equation}
\vspace{-0.4cm}

\noindent
where $0 \le {\theta _i} \le 1$, and $r_{i,j}$ denotes the number of RBs allocated by BS $i$ to UE $j$. Therefore, if the power consumption of a functional block (e.g., cooling) common to all sectors of the BSs is proportional to the number of sectors, the total energy consumption of BS $i$ at time $t$ is expressed by

\vspace{-0.3cm}
\begin{equation}
{E_i} = \left( {1 - {\eta _i}} \right){\theta _i}p_i^{\max } + {\eta _i}p_i^{\max },
\end{equation}
\vspace{-0.5cm}

\noindent
where $p_i^{\max }$ is the maximum transmit power of BS $i$, and expressed as $p_i^{\max } = {g_i}{p_i} + {h_i}$, where $g_i$ and $h_i$ are two constants, separately, and $p_i$ is the transmit power of BS $i$. $\eta_i$ satisfies the interval of $[0,1]$. If ${\eta _i} = 1$, we utilize a constant energy consumption model. If ${\eta _i} = 0$, a fully energy proportional model is used to measure the energy consumption of BS $i$. Otherwise, ${\eta _i} \in \left( {0,1} \right)$ is a non-energy proportional model.

\subsection{Communication Model}
For measuring the throughput of the downlink, $K$ UEs with communication requirements in the cell establish wireless access, and then eatablish the downlinks to $M$ BSs. We consider downlink communications using the wireless access mode of OFDM between the BSs and the UEs \cite{AbdelnasserA2016}. We assume that the UEs can obtain the accurate channel state information. The distance between BS $i$ and UE $j$ at time $t$ is formulated by ${d_{i,j}} = \sqrt {L_{\rm{B}}^2+{{\left( {{x_i} - {x_j}} \right)}^2} + {{\left( {{y_i} - {y_j}} \right)}^2} }$. Since the urban macro-cellular scenario is considered in this paper, we utilize a 3DUma line-of-sight transmission model to measure the downlink channel \cite{ZhangK2019}. Thus, the path loss between BS $i$ and UE $j$ at time $t$ can be calculated as 

\vspace{-0.2cm}
\begin{equation}
{l_{i,j}} = 28.0 + 22{\log _{10}}\left( {{d_{i,j}}} \right) + 20{\log _{10}}\left( {{f_{\rm{C}}}} \right),
\end{equation}
\vspace{-0.3cm}

\noindent
where $f_{\rm{C}}$ is the system frequency. The channel gain between BS $i$ and UE $j$ at time $t$ is given by

\vspace{-0.2cm}
\begin{equation}
{c_{i,j}} = 10 - 20\lg \left( {\cos \frac{{\pi {\beta _i}}}{{180}}} \right) - {l_{i,j}} - \alpha,
\end{equation}
\vspace{-0.3cm}

\noindent
where $\beta_i$ is the adjustment of the antenna angle for BS $i$, and $\alpha$ denotes the shadow attenuation. Due the target signal is from the serving BS to the UEs, the interference signals is that the signals observes from other BS $\varphi$, for $\varphi \in {\cal M}$, $\varphi \ne i$. When BS $\varphi$ uses the transmit power $p_{\varphi}$, the overall interference from other BSs to UE $j$ is equal to $\sum\nolimits_{\varphi  \in {{\cal M}},\varphi  \ne i} {{p_\varphi }} {c_{\varphi ,j}}$. Thus, the downlink throughput between BS $i$ and UE $j$ at time $t$ is written as 

\vspace{-0.3cm}
\begin{equation}
{R_{i,j}} = {r_{i,j}}{B^{\rm RB}}{\log _2}\left( {1 + \frac{{{p_i}{c_{i,j}}}}{{\sum\nolimits_{\varphi  \in {{\cal M}},\varphi  \ne i} {{p_\varphi }{c_{\varphi ,j}}}  + \sigma _0^2}}} \right),
\end{equation}
\vspace{-0.1cm}

\noindent
where $\sigma _0^2$ is the noise variance.

\subsection{First Packet Latency Model}
The first packet latency from the BSs to the UEs mainly
depends on its queuing latency and transmission latency. We assume that the BSs adopts the \emph{first arrival, first processing} to serve the packets of the UEs. The packets arrive and wait in the buffer of the BSs. Let ${\varpi _{j,n}}$ and $\varpi _{j,n}^{\max }$ denote the coding rate and maximum coding rate of UE $j$ at CQI $n$, respectively. We then design a binary variable to depict
whether to perform the resource scheduling from BS $i$ to UE $j$ at time $t$, i.e.,

\vspace{-0.3cm}
\begin{equation}
{\zeta _{i,j}} = \left\{ {\begin{array}{*{20}{l}}
		{\!\!\!1,}&{{\text{if BS }}i{\text{ performs resource scheduling}},}\\
		{\!\!\!0,}&{{\text{otherwise}}{\rm{,}}}
\end{array}} \right.
\end{equation}
\vspace{-0.2cm}

\noindent
where ${\zeta _{i,j}}=1$ represents ${\varpi _{j,n}} \le \varpi _{j,n}^{\max }$, and ${\zeta _{i,j}}=0$ denotes ${\varpi _{j,n}} > \varpi _{j,n}^{\max }$. Let $T_{i,j}^Q={T_{i,j}^{\rm out}-T_{i,j}^{\rm in}}$ denote the queuing latency of the in-out time difference latency for the packets, where ${T_{i,j}^{\rm out}}$ and ${T_{i,j}^{\rm in}}$ is the time at which the packets of UE $j$ enters and leaves the buffer of BS $i$, respectively. As such, the first packet latency between BS $i$ and UE $j$ at time $t$ can be formulated as follows

\vspace{-0.3cm}
\begin{equation}
{T_{i,j}} = T_{i,j}^{\rm out} - T_{i,j}^{\rm in} + \frac{{\tau_{\rm pkt}}}{{\zeta _{i,j}^{}{\varpi _{j,n}}{r_{i,j}}R_{}^{\rm RB}}},
\end{equation}
\vspace{-0.2cm}

\noindent
where $R^{RB}$ and $\tau_{\rm pkt}$ are the rate of each RB and the bits of first packet, respectively.

\subsection{Problem Formulation}
As for the energy-aware RAN scenario, each network objective corresponds to a utility function, and can be abstracted as a multi-objective optimization problem. It is necessary to find the maximization or minimization solution for all network objectives that match network state. Thus, we jointly consider the three objectives of the total energy consumption $E_i$, the downlink throughput $R_{i,j}$, and the first packet latency $T_{i,j}$, the optimization problem of multiple network objectives can be expressed in the following manner
 
\vspace{-0.4cm}
\begin{flalign}
	{\mathop {\min }\limits_{{E_i},{R_{i,j}},{T_{i,j}},{\xi _{i,j}}} }& {\left\{ {{E_i},\: -\sum\limits_{j = 1}^K {{\xi _{i,j}}{R_{i,j}},\: \sum\limits_{j = 1}^K {{\xi _{i,j}}{T_{i,j}}} } } \right\}}  \label{YY}\\
	{\rm s.t.} \ \ \ \ \ \ &{{p_i} \le p_i^{\max }},\forall i, \tag{\ref{YY}{a}} \label{YYa}\\
	&{\sum\limits_{j = 1}^K {{r_{i,j}}} } \le \frac{{{B_i}}}{{{B^{\rm RB}}}},\forall i,j, \tag{\ref{YY}{b}} \label{YYb}\\
	&{{p_i}{c_{i,j}} \ge {p_0},} \forall i,j, \tag{\ref{YY}{c}} \label{YYc}\\
	&{\sum\limits_{i = 1}^M {{E_i}} } \le {E^{\max }},\forall i, \tag{\ref{YY}{d}} \label{YYd}\\
	&{\sum\limits_{i = 1}^M {\sum\limits_{j = 1}^K {{\xi _{i,j}}{R_{i,j}} \ge {R^{\min }},\forall i,j,} } } \tag{\ref{YY}{e}} \label{YYe}\\
	&{\sum\limits_{i = 1}^M {\sum\limits_{j = 1}^K {{\xi _{i,j}}{T_{i,j}} \le {T^{\max }},\forall i,j,n,} } } \tag{\ref{YY}{f}} \label{YYf}\\
	&{{\xi _{i,j}},{\zeta _{i,j}} \in \left\{ {0,1} \right\},}\forall i,j, \tag{\ref{YY}{g}} \label{YYg}
\end{flalign}
\vspace{-0.5cm}

\noindent
where $p_i^\text{max}$ and $p_0$ denote the maximum transmit power of BS $i$ and the minimum received signal power of UE $j$, respectively. $E^\text{max}$, $R^\text{min}$, and $T^\text{max}$ is the maximum total energy consumption, the minimum downlink throughput, and the maximum first packet latency from $M$ BSs to $K$ UEs. In addition, the constraints (9a) and (9b) ensure the maximum transmit power and the maximum number of RBs allocated by BS $i$ to UE $j$, respectively. The constraint (9c) shows that the received signal power of UE $j$ must be larger than or equal to the minimum received signal power $p_0$. The constraints (9d), (9e), and (9f) are utilized to guarantee the performance requirements of the total energy consumption, the downlink throughput, and the first packet latency. Specially, the accurate value of $E^\text{max}$, $R^\text{min}$, and $T^\text{max}$ are derived from the settings of network intent, where a specific example named as ``\emph{Ensure that the total energy consumption is $\le$ 0.6KWh, the downlink throughput is $\ge$ 0.5Gbps, and the first packet latency is $\le$ 1ms}'' is provided in Section \ref{sectionIV-A}. The association constraints of the downlink throughput $R_{i,j}$ and the first packet latency $T_{i,j}$ are demonstrated in (9g).

\section{Energy-Aware Network Intent Representation, Transformation and Modeling}
\label{sectionIV}
We concentrate on a network intent in various representation formats in this section. We propose a method about the representation of intent ontology based on 3GPP template. Then, the network intent is transformed YAML to JSON format. Moreover, KAOS modeling language is introduced to describe network ontology within energy-aware RAN scenario. 

\begin{figure}[t]
	\centering
	\includegraphics[width=2.1in]{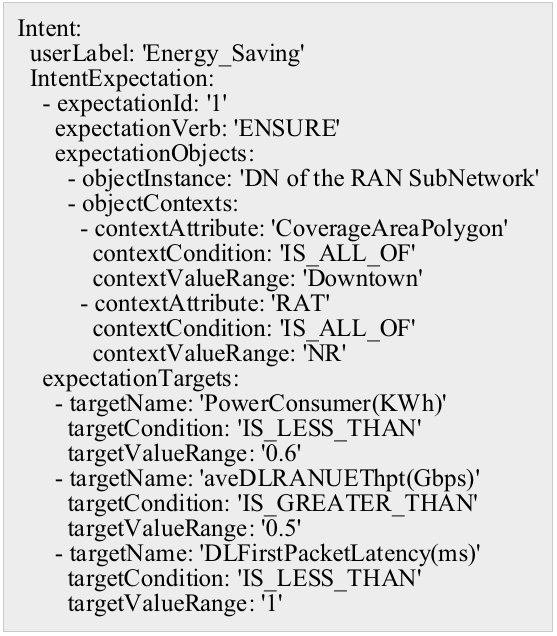}
	\caption{Example of network intent in YAML format.}
	\label{Fig_2_Intent_YAML}
	\vspace{-0.5cm}
\end{figure}

\vspace{-0.1cm}
\subsection{Intent Representation Using 3GPP Template}
\label{sectionIV-A}
According to the standard of 3GPP TS 28.312 \cite{StandardIntent}, we provide and characterize an example of a network intent in YAML format, as shown in Fig. \ref{Fig_2_Intent_YAML}. This example is named as ``\emph{Ensure that the total energy consumption is $\le$ 0.6KWh, the downlink throughput is $\ge$ 0.5Gbps, and the first packet latency is $\le$ 1ms}''. The intent representation by an intent template of 3GPP standard mainly includes two components, such as \emph{expectationObjects} and \emph{expectationTagets}.

\subsubsection{Intent Ontology Based on 3GPP Standard}
Based on 3GPP standard, each intent consists of many elements, which is called intent ontology. Formally, the representation framework of the intent ontology is defined in the following.

\begin{figure}[t]
	\centering
	\includegraphics[width=2.1in]{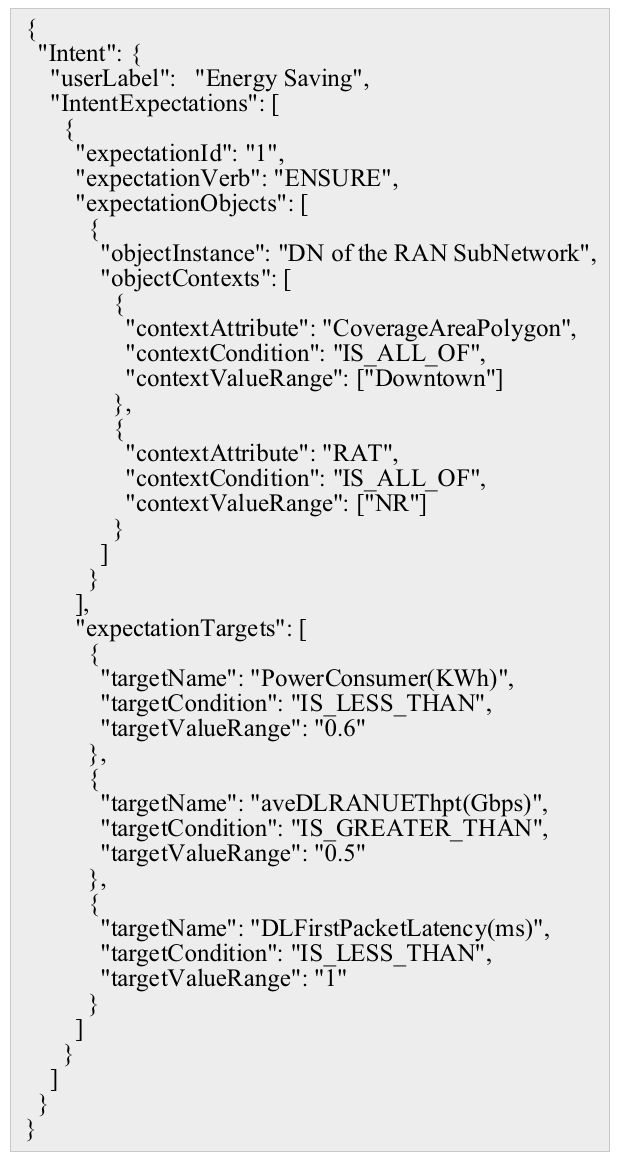}
	\caption{Example of intent transformation from YAML to JSON format.}
	\label{Fig_3_Intent_JSON}
	\vspace{-0.5cm}
\end{figure}

\noindent
\textbf{Definition 1.} \emph{(Intent Ontology):} An example of network intent called ``\emph{Ensure that the total energy consumption is $\le$ 0.6KWh, the downlink throughput is $\ge$ 0.5Gbps, and the first packet latency is $\le$ 1ms}'' can be formulated as a 4-tuple

\vspace{-0.25cm}
\begin{equation}
\begin{array}{c}
	\!\!\!\!\!\!\!\!\!\!\!\!\!\!\!\!\!\!\!\!\!\!\!\!\!\!\!\! \mathcal{IO} = \left\{ {userLabel,IntentExpectation,} \right.\\
\end{array}\nonumber
\end{equation}
\vspace{-0.8cm}
\begin{equation}
	\begin{array}{c}
		\left. {\ \ \ \ \ \ \ \ \ expectationObjects,expectationTargets} \right\}\!,
	\end{array} \!\!\!\!\!\!\!
\end{equation}
\vspace{-0.6cm}

\begin{itemize}
	\item \textit{userLabel}: This is used to state the intent expectation, i.e., ``\textit{Energy Saving}''.
	
	\item \textit{IntentExpectation}: Consisting of all expressed contents for the intent expectation. It mainly includes \textit{expectationId}, \textit{expectationVerb}, \textit{expectationObjects}, and \textit{expectationTargets}. Therein, \textit{expectationId} is the desired number of this intent expectation, i.e., ``$1$''. \textit{expectationVerb} is the characteristic of this intent expectation, and we denote it as ``\textit{ENSURE}''. \textit{expectationObjects} and \textit{expectationTargets} are two important components, representing the state of the BSs in the energy-aware RAN scenario and the target state that needs to be reached, respectively.

	\item \textit{expectationObjects}: The \textit{expectationObjects} stands for the objects of the intent expectation. There are \textit{objectInstance} and \textit{objectContexts}, denoted by a sample of the objects. To be more specific, \textit{objectInstance} is set to ``\textit{DN of the RAN SubNetwork}''. \textit{objectContexts} is included in detail as \textit{contextAttribute}, \textit{contextCondition}, and \textit{contextValueRange}. Therin, \textit{contextAttribute} shows the attribute of the object, set as ``\textit{CoverageAreaPolygon}'' and ``\textit{RAT}'', respectively. \textit{contextCondition} is the condition of the object, given as ``\textit{IS\_ALL\_OF}''. Let \textit{contextValueRange} denotes the value range of the object, and consists of ``\textit{Downtown}'' and ``\textit{NR}''.
	
	\item \textit{expectationTargets}: The \textit{expectationTargets} obtained by the targets of the intent expectation, and mainly includes \textit{targetName}, \textit{targetCondition}, and \textit{targetValueRange}. \textit{targetName} is the name involving three targets. The total energy consumption, the downlink throughput, and the first packet latency are expressed as ``\textit{PowerConsumer(KWh)}'', ``\textit{aveDLRANUEThpt(Gbps)}'', and ``\textit{DLFirstPacketLatency(ms)}'', respectively. The condition and the value range of ``\textit{PowerConsumer(KWh)}'' are ``\textit{IS\_LESS\_THAN}'' and ``\textit{0.6}'', separately. Similarly, two other targets are also represented to include the condition and the value range. ``\textit{aveDLRANUEThpt(Gbps)}'' and ``\textit{DLFirstPacketLatency(ms)}'' include ``\textit{IS\_GREATER\_THAN}'' and ``\textit{IS\_LESS\_THAN}'', ``\textit{0.5}'' and ``\textit{1}'', respectively. In particular, $E^{\rm max}$, $R^{\rm min}$, and $T^{\rm max}$ of constraints (9d), (9e), and (9f) are related to \textit{expectationTargets}, more specifically to \textit{targetValueRange} of three targets. In addition, the inequality relation of constraints (9d), (9e), and (9f) are denoted by \textit{targetCondition} of \textit{expectationTargets} for three targets, respectively.
	
\end{itemize}

\subsubsection{Intent Transformation Using JSON Format}
The network intent based on YAML format of 3GPP standard has many defects in actual software engineering applications. It is not enough to support dynamic and real-time update of this intent. Therefore, transforming the network intent from YAML to JSON format can better storage the data. We find that JSON format is more suitable for storing the data of network intent than YAML format. As illustrated in Fig. \ref{Fig_3_Intent_JSON}, we realize the transformation between YAML format and JSON format, that is, the standard intent template of YAML format is transformed into JSON format. Specifically, an example of network intent called ``\emph{Ensure that the total energy consumption is $\le$ 0.6KWh, the downlink throughput is $\ge$ 0.5Gbps, and the first packet latency is $\le$ 1ms}'' is defined and formulated again in the form of 4-tuple within $\mathcal{IO}$ of \textbf{Definition 1}.

\subsection{Intent Modeling Based on KAOS Language}
\begin{figure}[t]
	\centering
	\includegraphics[width=3.5in]{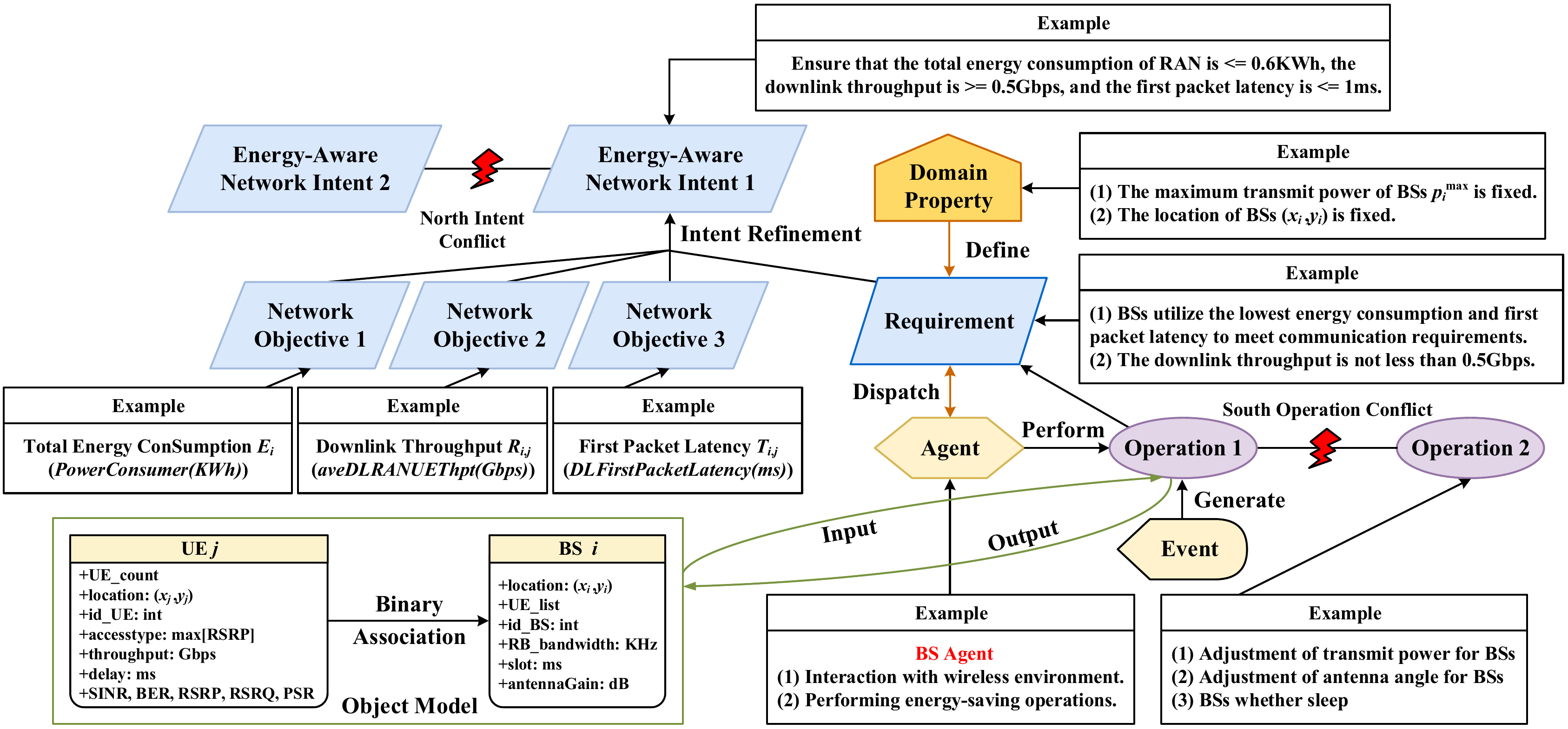}
	\caption{A modeling example of network intent using \textit{Knowledge} Acquisition in automated Specification language.}
	\label{Fig_5_KAOS_IntentModeling}
	\vspace{-0.5cm}
\end{figure}

The method of intent modeling based on KAOS language mainly depends on domain knowledge and manual annotation by experts. The experts first define the concepts and relationships of intent modeling, and then rely on multiple network objectives to refine and expand the network intent. Finally, a complete knowledge base related to intent modeling is constructed. As shown in Fig. \ref{Fig_5_KAOS_IntentModeling}, a modeling example of network intent based on KAOS language is considered. It mainly models the existing elements and network ontologies.

\noindent
\textbf{Definition 2.} \emph{(Network Ontology):} A modeling example of network intent called ``\emph{Ensure that the total energy consumption is $\le$ 0.6KWh, the downlink throughput is $\ge$ 0.5Gbps, and the first packet latency is $\le$ 1ms}'' using KAOS language is formulated as 6-tuple

\vspace{-0.5cm}
\begin{equation}
	\begin{array}{c}
		\mathcal{\! NO} \!=\! \left\{ {Objective,DomainProperty,RANRequirement} \right.\\
	\end{array}\nonumber
\end{equation}
\vspace{-0.8cm}
\begin{equation}
	\begin{array}{c}
		\left. {\ \ \  EnergySavingOP,BSAgent,ConflictRule} \right\}\!,
	\end{array}\!\!\!\!\!\!\!\!\!
\end{equation}
\vspace{-0.65cm}

\begin{itemize}
	\item \textit{Objective}: The example of network intent can be refined into three network objectives, such as the total energy consumption $E_i$, the downlink throughput $R_{i,j}$, and the first packet latency $T_{i,j}$.
	
	\item \textit{DomainProperty}: The domain properties of the considered intent include ``\textit{the maximum transmit power of BSs $p_i^{\rm max}$ is fixed}'', and ``\textit{the location of BSs $(x_i,y_i)$ is fixed}''.
	
	\item \textit{RANRequirement}: Two \textit{RANRequirements} (i.e., constraints (9d), (9e), and (9f)) are proposed based on \textit{DomainProperty}, specifically as ``\textit{BSs utilize the lowest energy consumption and first packet latency to meet communication requirements}'', and ``\textit{the downlink throughput is not less than 0.5Gbps}''.
	
	\item \textit{EnergySavingOP}: The BSs can perform three energy-saving operations, such as ``\textit{the adjustment of transmit power for BSs}'', ``\textit{the adjustment of antenna angle for BSs}'', and ``\textit{the BSs whether sleep}''.
	
	\item \textit{BSAgent}: \textit{BSAgent} is an active ontology in the process of the network intent realization. It can select and perform \textit{EnergySavingOP} according to current wireless environment, and collect the information feedback to make decisions. \textit{RANRequirement} is analyzed on the basis of the UEs' information to which the managed BSs are connected.
	
	\item \textit{ConflictRule}: The conflicts exist in ``\textit{north intent}'' and ``\textit{south operation}''. We predefine the priority, conflict level, and conflict rules among network objectives and among operations, respectively.
\end{itemize}

Based on $\mathcal{NO}$ in \textbf{Definition 2}, the network intent modeling process of KAOS language is to build a complete knowledge base related to the components of the energy-aware RAN. This knowledge base not only refines the architecture of the energy-aware RAN, but also provides a prior knowledge for the network intent decomposition mentioned in Section-\ref{Section-V}.

\section{Energy-Aware Network Intent Decomposition} \label{Section-V}
In this section, we propose a network intent decomposition technique to implement the decomposition logic of goal-subgoal-operation. Taking three network objectives and three types of energy-saving operations as examples, we then construct a \textit{Softgoal} Interdependency Graph (SIG)-based decomposition model of network intent.

\begin{figure}[t]
	\centering
	\includegraphics[width=2.9in]{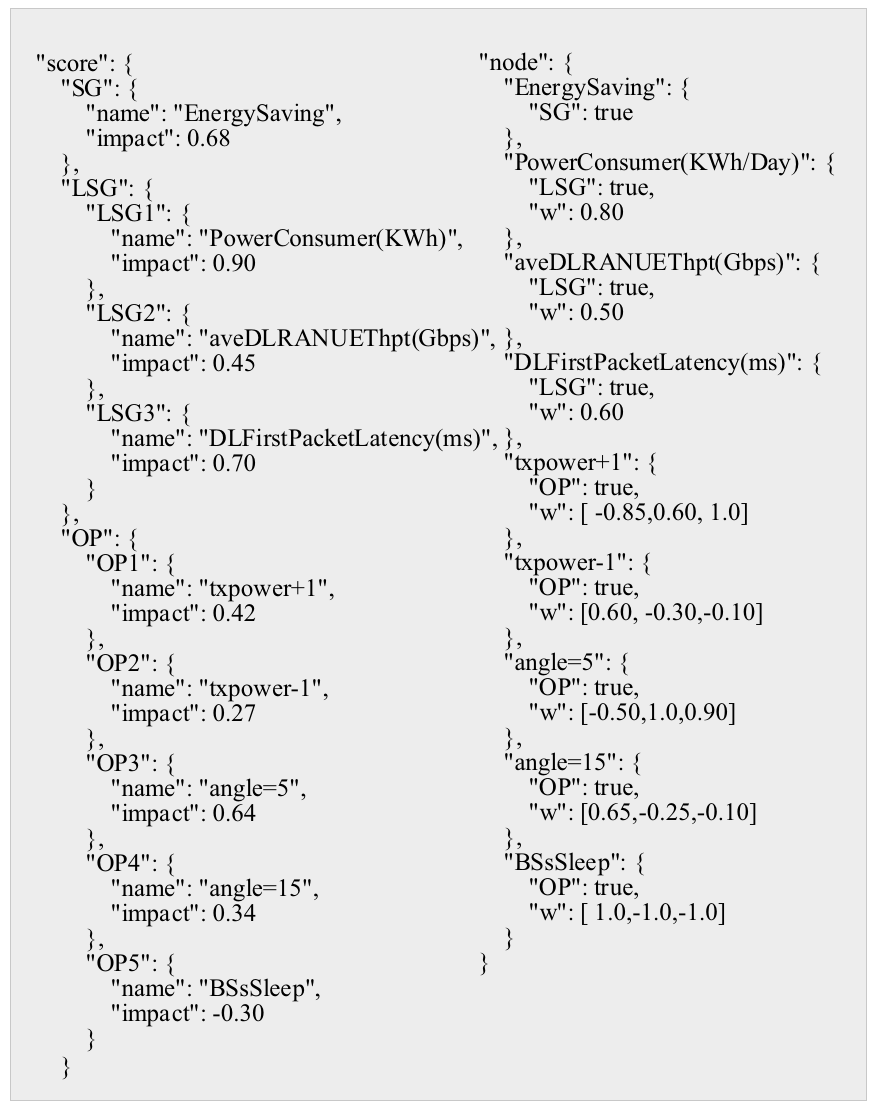}
	\caption{Example of \textit{Softgoal} Interdependency Graph decomposition in JSON format.}
	\label{Fig_7_SIG_JSON}
\vspace{-0.7cm}
\end{figure}

\subsection{Intent Decomposition Using SIG Model}
The SIG-based decomposition of network intent follows a top-down paradigm, starting from the decomposition of this intent into network objectives until energy-saving operations are defined and selected. We assume that the SIG decomposition model has been predefined, and that each energy-saving operation in down-top process quantifies the scores of the network intent and all network objectives, indicating its achievable granularity. Fig. \ref{Fig_7_SIG_JSON} provides a decomposition example of network intent based on the SIG in JSON format, and is defined in the following.

\noindent
\textbf{Definition 3.} \emph{(Intent Decomposition):} A decomposition example of network intent called ``\emph{Ensure that the total energy consumption is $\le$ 0.6KWh, the downlink throughput is $\ge$ 0.5Gbps, and the first packet latency is $\le$ 1ms}'' based on the SIG is formulated as a 5-tuple

\vspace{-0.2cm}
\begin{equation}
	\begin{array}{c}
		{{\cal S}{\cal I}{\cal G}} = \left\{ {{{\cal S}{\cal G}},{{\cal L}{\cal S}{\cal G}},{{\cal O}{\cal P}},{\cal W},{\cal S}} \right\},\\
	\end{array}
\end{equation}
\vspace{-0.6cm}

\begin{itemize}
	\item \textit{Energy-aware network intent} $\mathcal{SG}$: There have an network intent, decribed as $\mathcal{SG}$. 
	
	\item \textit{Network objectives set} $\mathcal{LSG}$: The network intent $\mathcal{SG}$ is decomposed and refined into three network objectives, such as ``\textit{Total Energy Consumption} $E_i$'', ``\textit{Downlink Throughput} $R_{i,j}$, and ``\textit{First Packet Latency} $T_{i,j}$, denoted as a set $\mathcal{LSG}$.
	
	\item \textit{Energy-saving OPs set} $\mathcal{OP}$: To achieve different network objectives, all energy-saving operations can be divided into ``\textit{Transmit Power of BSs $+1dBm$}'', ``\textit{Transmit Power of BSs $-1dBm$}'', ``\textit{Antenna Angle of BSs $={5^\circ}$}'', ``\textit{Antenna Angle of BSs $={15^\circ}$}'', and ``BSs Sleep'', which can be expressed as a set $\mathcal{OP}$.
	
	\item \textit{Weights} $\mathcal{W}$: As shown in Fig. \ref{Fig_7_SIG_JSON}, the weights between the network intent $\mathcal{SG}$ and each network objective are initially set as $\left\{ {0.80,0.50,0.60} \right\}$. In addition, the weights between five energy-saving OPs and each network objective are initially set as $\left\{ {-0.85,0.60,1.0} \right\}$, $\left\{ {0.60,-0.30,-0.10} \right\}$, $\left\{ {-0.50,1.0,0.90} \right\}$, $\left\{ {0.65,-0.25,-0.10} \right\}$, and $\left\{ {1.0,-1.0,-1.0} \right\}$.
	
	\item \textit{Scores} $\mathcal{S}$: It is readily calculated that the scores of five energy-saving OPs are $\left\{ {{\rm{0}}{\rm{.42,0}}{\rm{.27,0}}{\rm{.64,0}}{\rm{.34,-0}}{\rm{.30}}} \right\}$, the scores of three network objectives are $\left\{ {{\rm{0}}{\rm{.90,0}}{\rm{.45,0}}{\rm{.70}}} \right\}$, and the scores of the softgoal for the network intent $\mathcal{SG}$ are ``$0.68$'' \cite{ScheidEJ2017}.
\end{itemize}

\begin{figure}[t]
	\centering
	\includegraphics[width=3.5in]{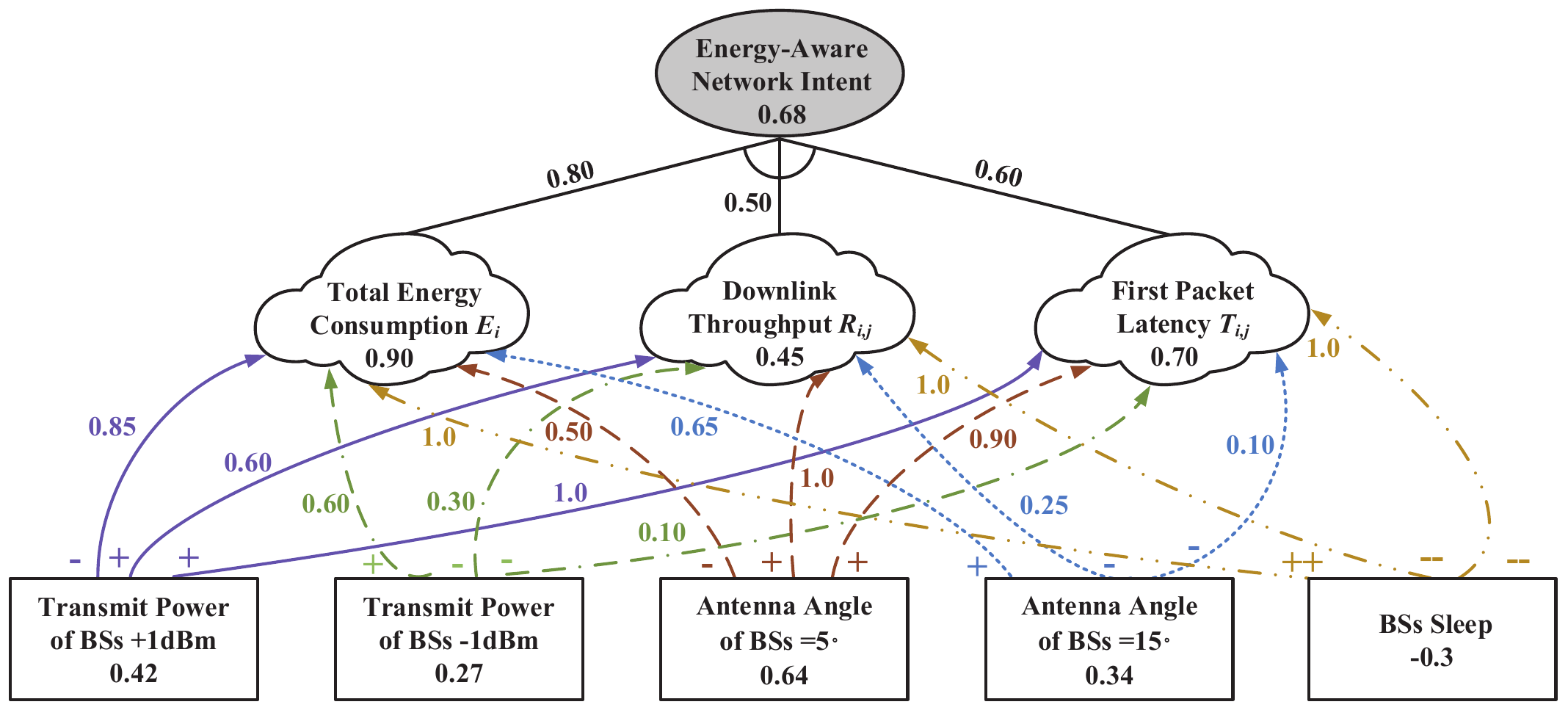}
	\caption{A decomposition example of network intent based on \textit{Softgoal} Interdependency Graph.}
	\label{Fig_8_SIG_IntentDecomposition}
	\vspace{-0.7cm}
\end{figure}

Through the above analysis, a graphical decomposition example of network intent $\mathcal{SG}$ is illustrated in Fig. \ref{Fig_8_SIG_IntentDecomposition}, where the SIG characterizes each tuple in \textbf{Definition 3} $\mathcal{SIG}$. The scores of the softgoal $S^{\rm SG}$ reflects the satisfiability degree of the network intent $\mathcal{SG}$. The domain expert manually sets the minimum score $S^{\rm th}$ when the network intent $\mathcal{SG}$ can be satisfied. Let us introduce a binary variable as follows to describe the association relationship between the scores of the softgoal $S^{\rm SG}$ and the minimum score $S^{\rm th}$, i.e., 

\vspace{-0.4cm}
\begin{equation}
{s^{\rm sc}} = \left\{ {\begin{array}{*{20}{l}}
		{\!\!1,}&{{\text{if network intent }}{{\cal S}}{{{\cal G}}}{\text{ is satisfied}},}\\
		{\!\!0,}&{{\text{otherwise}},}
\end{array}} \right.
\end{equation}
\vspace{-0.25cm}

\noindent
where ${s^{\rm sc}}=1$ represents ${S^{\rm SG}} \ge S^{\rm th}$, which indicates that the network intent ${{\cal S}}{{{\cal G}}}$ is satisfied. If ${s^{\rm sc}}=0$ denotes ${S^{\rm SG}} < S^{\rm th}$, which demonstates that the network intent ${{\cal S}}{{{\cal G}}}$ is not satisfied.

\subsection{Intent Decomposition Algorithm Design}
For the network intent decomposition process $\mathcal{SG}$, illustrating the intent ontology, network ontology, and decomposition logic is important. Inspired by the existence of the conflicts among network objectives \cite{ZhangJ2021}, we then analyze the conflicts in the decomposition process $\mathcal{SG}$ to refine the output network objectives set $\mathcal{LSG}$ and energy-saving OPs set $\mathcal{OP}$.

\noindent
\textbf{Definition 4.} \emph{(Conflict Analysis):} Given a network intent decomposition model $\mathcal{SIG}$, there are some conflicts among network objectives, which forms a conflict set $\mathcal{EC}$.
 
As depicted in Fig. \ref{Fig_8_SIG_IntentDecomposition}, we find that there exists two conflicts among three network objectives in $\mathcal{SG}$. From Fig. \ref{Fig_3_Intent_JSON}, the \textit{targetCondition} of \textit{expectationTargets} ``\textit{PowerConsumer(KWh)}'' and ``\textit{aveDLRANUEThpt(Gbps)}'' are ``\textit{IS\_LESS\_THAN}'' and ``\textit{IS\_GREATER\_THAN}'', respectively. The condition of the two \textit{expectationTargets} are opposite, indicating a conflict in the implementation process. Similarly, there is also a conflict between the \textit{expectationTargets} ``\textit{aveDLRANUEThpt(Gbps)}'' and ``\textit{DLFirstPacketLatency(ms)}'' due to different \textit{targetCondition}.

\begin{algorithm}   
	\caption{Energy-Aware Network Intent Decomposition}
	\begin{algorithmic}[1]
		\Require A network intent of YAML format in \textbf{Definition 1}.
		\Ensure Energy-saving operations set ${\mathcal{OP}}$ of JSON format in \textbf{Definition 3} ${\mathcal{SIG}}$.
		\Function {YAMLtoJSON}{$Intent\  Ontology$}   
		\If {$yaml\_data \text{\ is not empty}$}  
		\State $parsed\_data \gets \text{parse YAML  with} \ yaml\_data$  
		\State $json\_data \gets \text{covert} \ parsed\_data \ \text{to JSON}$ 
		\EndIf   
		\State \Return{$json\_data \ \mathcal{JSON}$}
		\EndFunction  
		\Function {KAOSModeling}{$Network\  Ontology$}
		\If {$network\_ontology \text{\ is added}$}  
		\State $Objective \gets \text{three network objectives}$  
		\State $DomainProperty \gets p_i^{\rm max} \ \text{and} \ (x_i,y_i) \ \text{is fixed} $ 
		\State $RANRequirement \gets  \text{(9d),\ (9e), \ and \ (9f)}$
		\State $EnergySavingOP \gets  \text{three operations}$ 
		\State $BSAgent \gets  \text{perform} \ EnergySavingOP$
		\State $ConflictRule \gets intents \  \text{and} \ operations$  
		\EndIf  
		\State \Return{$modeling\_result \ \text{in \textbf{Definition 2}} \ \mathcal{NO}$}  
		\EndFunction
		\Function{SIGDecomposition}{$Intent \ Decomposition$}   
		\State $\text{\textbf{Set}} \ \mathcal{SG},\mathcal{LSG}, \mathcal{OP},\mathcal{W} \gets \text{\textbf{Definition 3}} \ \mathcal{SIG}$
		\State $\text{\textbf{Calculate} the scores of softgoal, \textit{LSGs}, and \textit{OPs}}$
		\While{$\text{calculate} \ Scores \ \mathcal{S}$}  
		\If{$\text{load} \ Weights \ \mathcal{W}$}  
		\State $\text{Scores of} \ \mathcal{OP}, \ \mathcal{LSG}, \ \mathcal{SG}$    
		\EndIf  
		\EndWhile   
		\For{$\text{identify} \ conflict \ set \ \mathcal{EC}$}  
		\State $\text{Load} \  ConflictRule \ \text{of \textbf{Definition 2}} \  \mathcal{NO}$ 
		\EndFor  
		\State \Return{$\text{\textit{OPs} set} \ \mathcal{OP} \ \text{in \textbf{Definition 3}} \ \mathcal{SIG}$}   
		\EndFunction  
		
	\end{algorithmic}  
\end{algorithm} 

The procedure of the network intent $\mathcal{SIG}$ decomposition algorithm is summarized in \textbf{Algorithm 1}. To improve the overall performance of achieving an network intent $\mathcal{SIG}$ decomposition, constructing a decomposition model needs to be efficiently conducted via the representation of the intent ontology, the network ontology, and the decomposition logic. By introducing the intent ontology based on 3GPP template, we propose an intent representation and transformation function for converting YAML to JSON format, which is expressed as \textbf{function} $\text{YAMLtoJSON}(Intent\ Ontology)$ in \textbf{Algorithm 1}. For tractability, let us utilize KAOS language to denote the network ontology of \textbf{Definition 2} $\mathcal{NO}$ in 6-tuple, which is represented as \textbf{function} $\text{KAOSModeling}(Network\ Ontology)$ of \textbf{Algorithm 1}. To obtain the energy-saving OPs set $\mathcal{OP}$ in \textbf{Definition 3} $\mathcal{SIG}$, the scores calculation of $\mathcal{SG}$, three network objectives, and five energy-saving OPs, and the identification of the conflicts $\mathcal{EC}$ for \textbf{Definition 4} are considered, which is denoted as \textbf{function} $\text{SIGDecomposition}(Intent\ Decomposition)$. Furthermore, the SIG decomposition model of Fig. \ref{Fig_8_SIG_IntentDecomposition} implemented by codes can also be derived.

\section{Simulation Results}
\label{sectionVI}
In this section, we evaluate the performance of our proposed intent decomposition scheme in an energy-aware RAN scenario. We first initialize the parameters of the energy-aware RAN scenario, and analyze the initial decomposition performance of \textbf{Algorithm 1} to output the combination of energy-saving operation sets. Then, we combine a deep Q-network (DQN) algorithm, and the BSs perform offline training based on the combination of energy-saving operation set to select the optimal energy-saving operation. Moreover, we validate the performance gain of the DQN-assisted intent optimization scheme and two baselines. 
 
\vspace{-0.1cm} 
\subsection{Parameter Setting and Performance Analysis}

In detail, we consider a urban macro BSs model of 3GPP TR38.912 \cite{StandardMacro} consisting of $M=40$ BSs and $K=320$ UEs whose moving speeds follow a Gaussian distribution $|{\cal N}(3, 1)|{\rm m{s^{ -1}}}$. Each BS with a fixed altitude $L_{\rm B}=25{\rm m}$ uses $M_{\rm T}=1$ transmission antenna to coverage the UEs with $K_{\rm R}=1$ receiving antenna. The system bandwidth of each BS and the bandwidth of each RB are set as $B_i \! =\!
\left\{ {10,20,40,100} \right\}\text{MHz}$ and $B_{RB}=180\text{KHz}$, respectively. The service flow of each UE follows a Poisson distribution with service arrival rate $\chi_j \! =\! \left\{ {1,2,4,8} \right\}$. Regarding energy consumption model, unless otherwise noted, we set the transmit power of each BS as $p_i \! =\! \left\{ {50,51,52,53} \right\}\text{dBm}$, and two constants are $g_i=21.45$ and $h_i=354.44$, separately. For the communication model, the system frequency is set to $3.5\text{GHz}$, and the antenna angle of each BS can be selected in $\left\{ {{0^ \circ },{5^ \circ },{{15}^ \circ }} \right\}$. We set the shadow attenuation $\alpha$ and the noise variance $\sigma_0^2$ to be $[-15,15]\text{dB}$ and $-174\text{dBm/HZ}$, respectively. The minimum received signal power of each UE is set to $p_0=-5.1\text{dBm}$. The bits of first packet $\tau_{pkt}$ is equal to $320\text{bits}$.

\begin{figure}[t]
	\centering
	\includegraphics[width=3.3in]{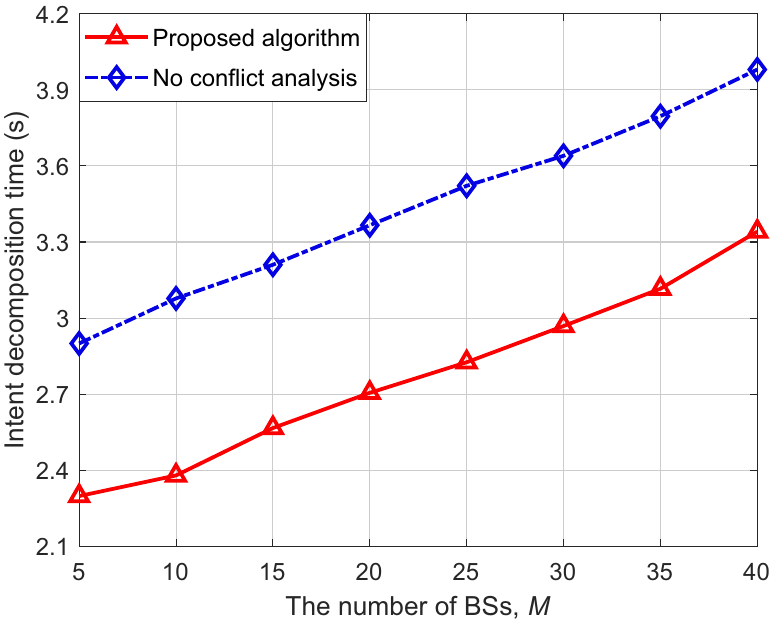}
	\caption{The initial decomposition time compared to the number of BSs between \textbf{Algorithm 1} and the baseline without the conflict analysis of \textbf{Definition 4}.}
	\label{Simulation_Results_1.eps}
	\vspace{-0.3cm}
\end{figure}

As shown in Fig. \ref{Simulation_Results_1.eps}, we validate the performance of \textbf{Algorithm 1}. Results in Fig. \ref{Simulation_Results_1.eps} show the comparsion of network intent decomposition time between the proposed algorithm and the baseline without the conflict analysis of \textbf{Definition 4}, versus the number of BSs $M$. As expected, we can find that the network intent decomposition time improve significantly for all the curves by increasing the value of $M$. This is attributed to the fact that more BSs are required for energy-saving, which requires more times of intent representation of \textbf{Definition 1} $\mathcal{IO}$, intent modeling \textbf{Definition 2} $\mathcal{NO}$, and intent decomposition \textbf{Definition 3} $\mathcal{SIG}$, leading to a dramatic growth of intent decomposition time for this energy-aware RAN scenario. As depicted in Fig. \ref{Simulation_Results_1.eps}, \textbf{Algorithm 1} outperforms the baseline in terms of the network intent decomposition time. This is mainly because the conflict analysis of \textbf{Definition 4} is not considered, the balance among network objectives will be broken, and the network intent cannot be decomposed in real-time for each BS, and thus, less intent decomposition time is obtained.

\begin{figure}
	\centering
	\subfigure[Proposed DQN-assisted scheme.]{
		\begin{minipage}[b]{0.382\textwidth}
			\includegraphics[width=1\textwidth]{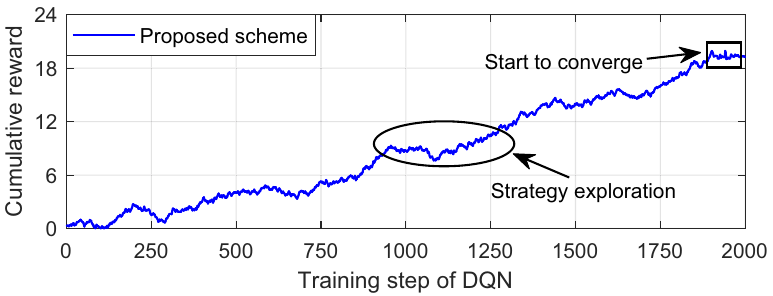}
		\end{minipage}
		\label{Simulation_Results_4_a}
	}
	\subfigure[No conflict analysis of \textbf{Definition 4} DQN-assisted scheme.]{
		\begin{minipage}[b]{0.382\textwidth}
			\includegraphics[width=1\textwidth]{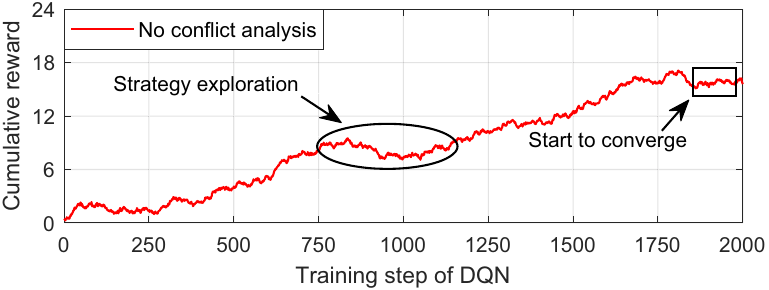}
		\end{minipage}
		\label{Simulation_Results_4_b}
	}
	\caption{The effect of the number of training steps for the DQN on the value of cumulative reward $r_i(t)$ for Eq. (\ref{eq20}) with the comparison of two schemes.}
	\label{Simulation_Results_4}
	\vspace{-0.3cm} 
\end{figure}

\subsection{DQN-Assisted Intent Optimization Scheme}
We consider a deep Q-network (DQN) algorithm with three neural networks to approximate the Q-function, so as to evaluate the decomposition performance of the network intent $\mathcal{SG}$. At time step $t$, \textit{BSAgent} $i$ in \textbf{Definition 2} $\mathcal{NO}$ returns the current state $s_i(t)$, and then selects the optimal energy-saving OPs ${\textbf{OP}_i^\textbf{*}}$. Subsequently, the state is transitioned from $s_i(t)$ to $s_i(t+1)$, and \textit{BSAgent} $i$ receives and stores reward $r_i(t)$ in replay memeory. By restricting the contribution of three network objectives to overall reward, the cumulative reward of \textit{BSAgent} $i$ at time step $t$ can be given by

\begin{figure*}[htb]
	\vspace{-0.3cm}
	\centering
	\subfigcapskip=-6.1pt
	\subfigure[]{
		\begin{centering}
			\includegraphics[scale=0.4]{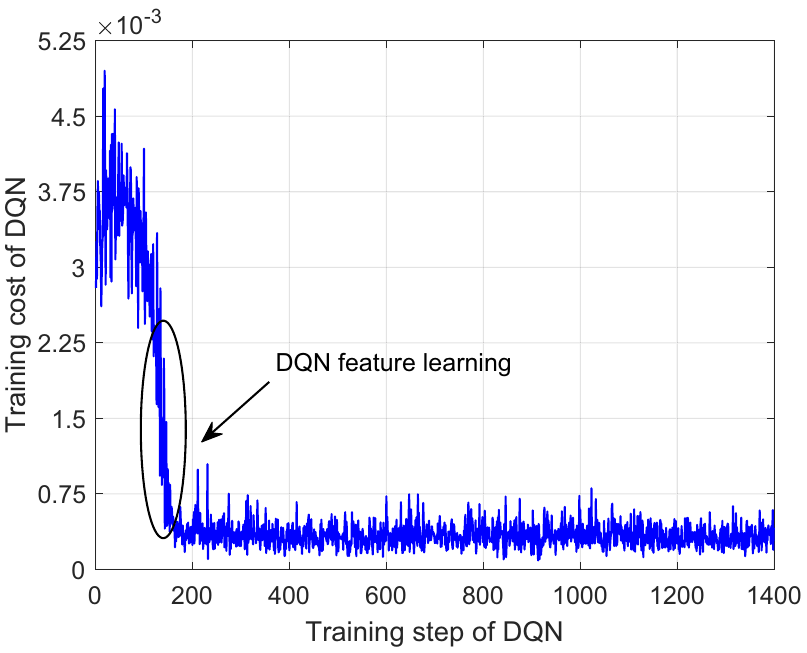}
		\end{centering}\hspace{-4mm}
		\label{Simulation_Results_2_a}}
	\subfigure[]{
		\begin{centering}
			\includegraphics[scale=0.4]{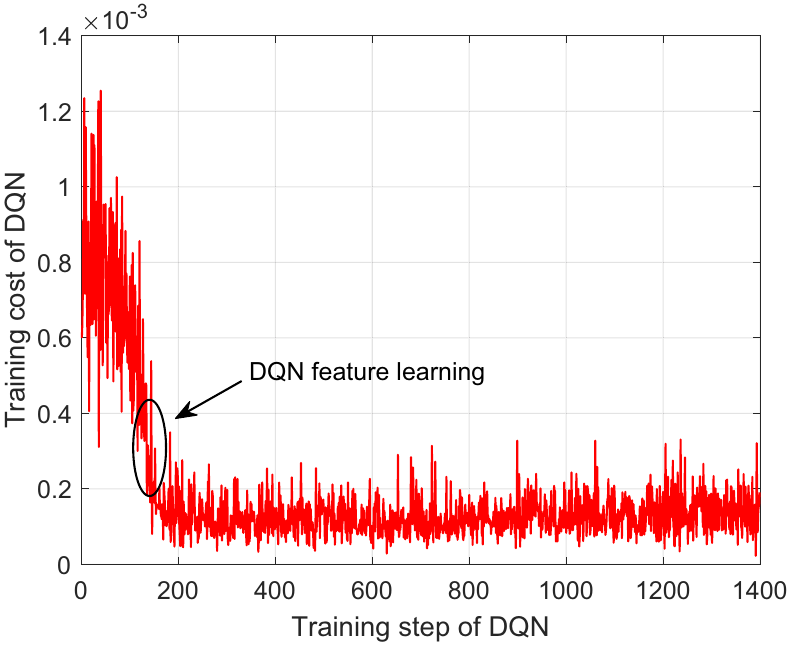}
		\end{centering}\hspace{-4mm}
		\label{Simulation_Results_2_b}}
	\subfigure[]{
		\begin{centering}
			\includegraphics[scale=0.4]{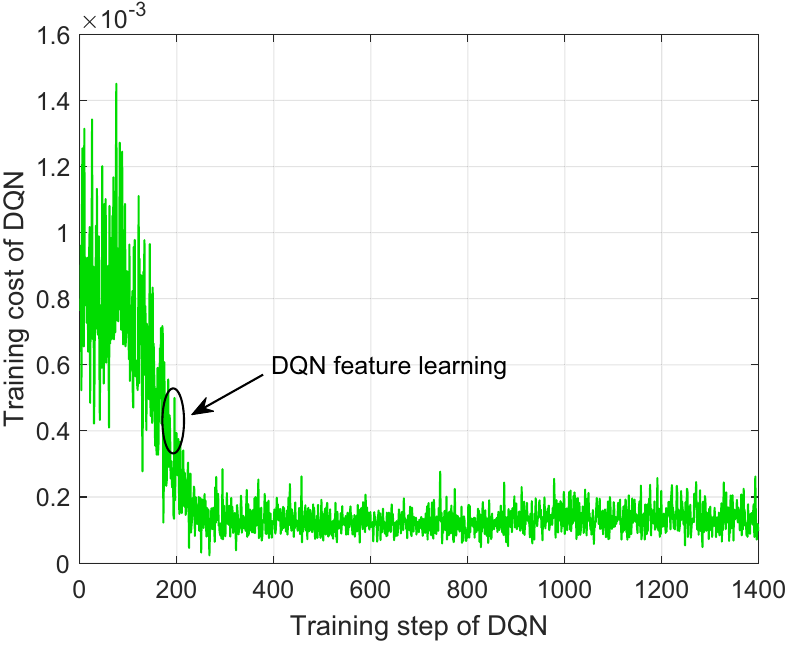}
		\end{centering}
		\label{Simulation_Results_2_c}}
	\vspace{-0.3cm}
	\caption{The training cost of DQN versus: (a) the learning rate of DQN is set as $0.01$; (b) the learning rate of DQN is set as $0.005$; (c) the learning rate of DQN is set as $0.001$, respectively, under $M=40$ BSs and $K=320$ UEs.}
	\label{Simulation_Results_2}
	\vspace{-0.48cm}
\end{figure*}

\begin{figure*}[htb]
	\centering
	\subfigcapskip=-6.1pt
	\subfigure[]{
		\begin{centering}
			\includegraphics[scale=0.4]{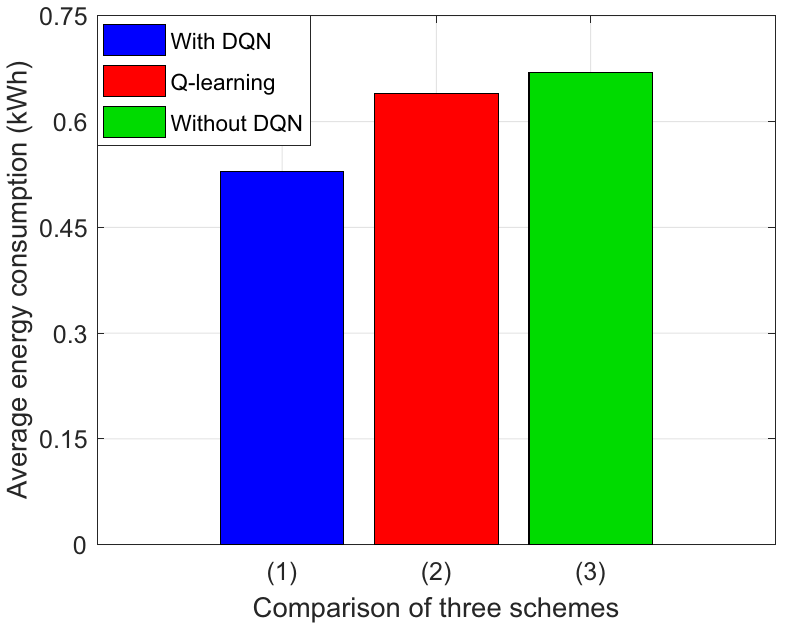}
		\end{centering}\hspace{-4mm}
		\label{Simulation_Results_3_a}}
	\subfigure[]{
		\begin{centering}
			\includegraphics[scale=0.4]{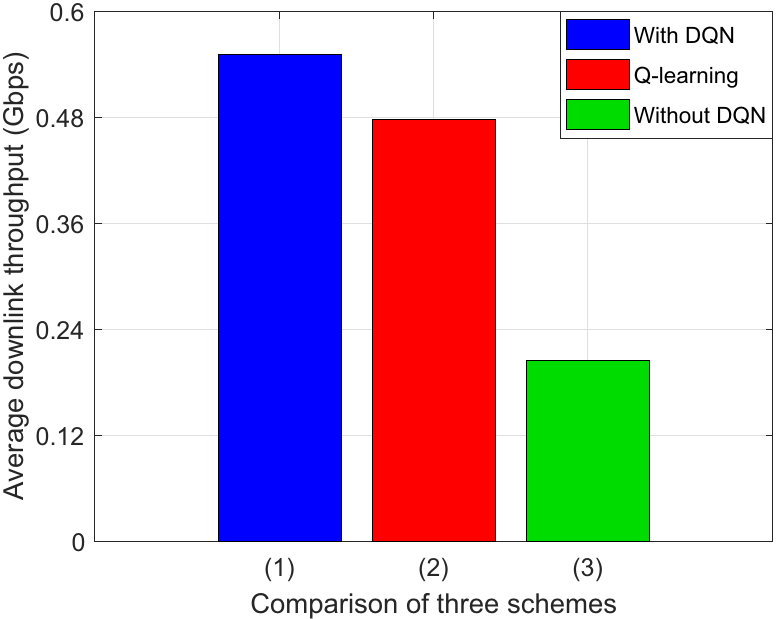}
		\end{centering}\hspace{-4mm}
		\label{Simulation_Results_3_b}}
	\subfigure[]{
		\begin{centering}
			\includegraphics[scale=0.4]{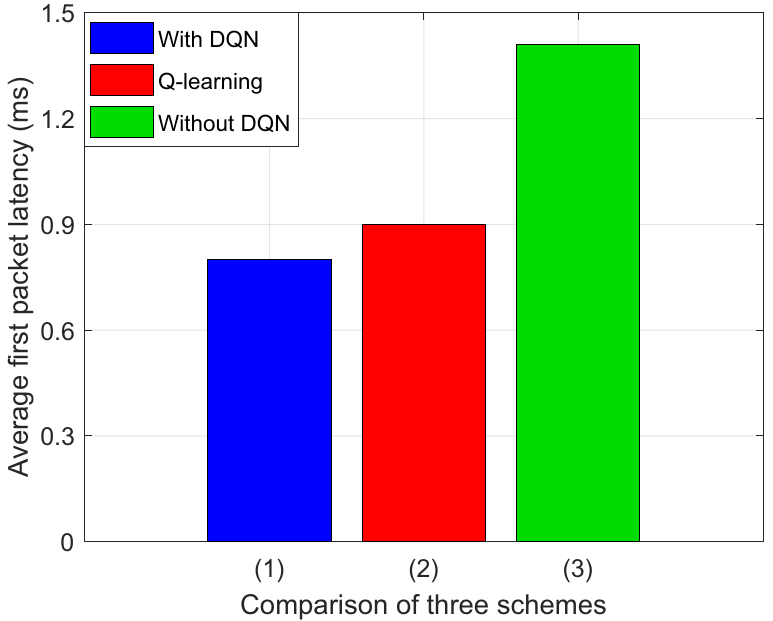}
		\end{centering}
		\label{Simulation_Results_3_c}}
	\vspace{-0.3cm}
	\caption{The performance of three network objectives in the average state versus the proposed scheme and two baselines with Q-learning-assisted scheme\cite{WilhelmiF2017} and no DQN-assisted scheme, respectively.}
	\label{Simulation_Results_3}
	\vspace{-0.48cm}
\end{figure*}

\vspace{-0.4cm}
\begin{equation}
{r_i}\left( t \right) = \frac{{{\delta _1}\left( {\overline {{R_i}}  - {R^{\min }}} \right)}}{{{R^{\max }} - {R^{\min }}}} - \frac{{{\delta _2}\left( {\overline {{E_i}}  - {E^{\min }}} \right)}}{{{E^{\max }} - {E^{\min }}}} - \frac{{{\delta _3}\left( {\overline {{T_i}}  - {T^{\min }}} \right)}}{{{T^{\max }} - {T^{\min }}}},\label{eq20}
\end{equation}
\vspace{-0.4cm}

\noindent
where $\delta_1$, $\delta_2$, and $\delta_3$ are non-negative weighting factors
announced from \textit{BSAgent} $i$, indicating its desire to tradeoff the downlink throughput, the total energy consumption, and the first packet latency. $\overline {{R_i}}  = \sum\nolimits_{j \in {{\cal K}}} {{R_{i,j}}/K}$, $\overline {{E_i}}  = \sum\nolimits_{j \in {{\cal K}}} {{E_{i,j}}/K}$, and $\overline {{T_i}}  = \sum\nolimits_{j \in {{\cal K}}} {{T_{i,j}}/K}$ denote the average value of the downlink throughput, the total energy consumption, and the first packet latency for \textit{BSAgent} $i$ at time step $t$, respectively. $R^{\rm max}$, $E^{\rm min}$, and $T^{\rm min}$ is the maximum downlink throughput, the minimum toal energy consumption, and the minimum first packet latency, respectively. The non-negative weighting factors are set to be $\delta_1=0.8$, $\delta_2=0.6$, and $\delta_3=0.2$ in Eq. (\ref{eq20}), respectively. Furthermore, we utilize a DQN model with a $3$-layer neural network and $64$ neurons. Both the reward discount factor and the greedy strategy index are set to $0.7$. For the target network, the number of update iterations is set to $100$. During the offline training process of DQN model, the cache of the experience replay pool for the update of network parameters is set to $3000$ experiences. The offline training process needs to iterate $1000$ time steps per episode, and each time step is $100$ ms.

In Fig. \ref{Simulation_Results_4}, we illustrate the value of cumulative reward $r_i(t)$ for Eq. (\ref{eq20}) versus the number of training steps of the DQN for the proposed scheme and the baseline with no conflict analysis of \textbf{Definition 4}. We easily observe that the increase of the number of training steps drastically affects the cumulative reward of the DQN $r_i(t)$ achieved by the proposed framework and the designed baseline in both Fig. \ref{Simulation_Results_4_a} and Fig. \ref{Simulation_Results_4_b}. Furthermore, it is evident that the proposed scheme outperforms the baseline in terms of higher cumulative reward $r_i(t)$. This is due to that no conflict analysis DQN-assisted scheme increases the space of energy-saving operations, and thereupon reduces the cumulative reward of the DQN training process $r_i(t)$. Moreover, we also can find that two subfigures of Fig. \ref{Simulation_Results_4} converge within the training step interval of $[1800,2000]$. As the value of cumulative reward $r_i(t)$ markedly growth, the trend of the curves may appear to increase and decrease at times. To explain, this trend suggests that the DQN has been undergoing strategy exploration, and the cumulative reward $r_i(t)$ may even be close to $0$ at the beginning of training (e.g., the training range of $[0,100]$).

Results in Fig. \ref{Simulation_Results_2} show the comparison of the training cost of the DQN algorithm with the learning rate of $0.01$, $0.005$, and $0.001$, respectively, versus the number of the training steps for the DQN algorithm from $0$ to $1400$. In Fig. \ref{Simulation_Results_2_a}, we plot the training cost the DQN algorithm with a learning rate of $0.01$ in the DQN-assisted intent optimization process. Overall, the curve follows a downward trend with training steps ranging from $0$ to $1400$. From Fig. \ref{Simulation_Results_2_a}, it can be easily observed that the training cost sharply decreases during the training steps in the range of $[120,180]$. This can be explained as follows: 1) A larger value of learning rate indicates that the features can be learned very quickly; and 2) the faster learning comes at a greater cost (i.e., the training cost). Similarly, the trend of two curves in Fig. \ref{Simulation_Results_2_b} with the learning rate of $0.005$ and Fig. \ref{Simulation_Results_2_c} with the learning rate of $0.001$ are still both downward. Looking globally at Fig. \ref{Simulation_Results_2}, despite higher learning rates learn the features faster, they incur more training costs. Meanwhile, compared to Fig. \ref{Simulation_Results_2_b}, lower learning rates in Fig. \ref{Simulation_Results_2_c} may tend to incur higher training costs when the training is just beginning, but later training will be more stable. This is not much surprising since a lower learning rate promotes a more stable convergence process.

Finally, Fig. \ref{Simulation_Results_3} presents the comparsion of the performance of three netowrk objectives in the average state between the proposed scheme and two baselines with Q-learning-assisted scheme \cite{WilhelmiF2017} and no DQN-assisted scheme, respectively. In Fig. \ref{Simulation_Results_3_a}, we illustrate the performance of average total energy consumption $\overline {{E_i}}$ between the proposed scheme and two baselines. It is evident from this figure that the average total energy consumption $\overline {{E_i}}$ markedly decreases with the continuous evolution of the proposed scheme. This is due to the fact that as the spatial dimension of energy-saving operation becomes larger, Q-learning slows down the learning rate and reduces the generalization ability, and therefore obtains higher average total energy consumption $\overline {{E_i}}$. We can also observe that no DQN-assisted scheme further lead to the continued decline of the average total energy consumption $\overline {{E_i}}$. Furthermore, as observed from Fig. \ref{Simulation_Results_3_b}, it compares the average downlink throughput $\overline {{R_i}}$ between the proposed scheme and two baselines. As expected, if the proposed scheme is adopted, the average downlink throughput $\overline {{R_i}}$ can be improved better accordingly no matter which baseline is utilized. Fig. \ref{Simulation_Results_3_c} illustrates the average first packet delay $\overline {{T_i}}$ versus the proposed framework and two baselines. For the same reason as in Fig. \ref{Simulation_Results_3_b}, the proposed framework can obtain better performance of average first packet delay $\overline {{T_i}}$ specifically. Such result suggests that DQN-assisted scheme will demonstrate better performance in the explosion of spatial dimension of energy-saving operations.

\section{Conclusions}
\label{sectionVII}
In this paper, we explored the network intent decomposition and optimization problem in the energy-aware radio access network scenario. To represent the network intent, we formulated the intent ontology from the YAML-based 3GPP template. We converted the network intent from YAML to JSON format for better data analysis. The network ontology set of the RAN was modeled using the Knowledge Acquisition in automated Specification modeling language. Relying on the intent and network ontologies, we performed the network intent decomposition based on the Softgoal Interdependency Graph approach. A network intent decomposition algorithm was proposed with conflict analysis among network objectives, and a composition of energy-saving operations was produced. Simulation results demonstrated the effectiveness and practicality of our proposed algorithm. A deep Q-network-assisted intent optimization scheme was designed, and we also demonstrated the performance gain of this scheme on each network objective and the reward function. This insight will shed light on applying practical network intent decomposition with enhanced effectiveness and accuracy.

\bibliographystyle{ieeetr}

\end{document}